\documentclass[a4paper]{article}

\usepackage{amsmath}
\usepackage{amssymb}
\usepackage{graphicx}
\usepackage{psfrag}
\usepackage[all]{xy}

\newtheorem{mththm}{Metatheorem}

\title{Turning the Liar Paradox into a Metatheorem of Basic Logic}

\author{P. A. Zizzi}

\date{\ }

\begin{document}

\maketitle

\vspace{-1cm}

\begin{center}
\emph{Dipartimento di Matematica Pura ed Applicata\\
Via Trieste, 63 - 35121 Padova, Italy\\
e-mail: zizzi@math.unipd.it}
\end{center}

\vspace{0.5cm}

{\abstract
We show that self-reference can be formalized in Basic logic
by means of the new connective: $@$, called ``entanglement''. In
fact, the property of non-idempotence of the connective $@$ is a
metatheorem, which states that a self-ontangled sentence loses
its own identity. This prevents having self-referential paradoxes
in the corresponding metalanguage. In this context, we introduce a generalized definition of self-reference, which is needed to deal with the multiplicative connectives of substructural logics.}

\section{Introduction}

Since Epimenides, the Greek philosopher who lived about 600 BC,
the Liar paradox: ``All Creteans are liars'' (in its modern version: ``This sentence is false'') remained ``unsolved''.

In our opinion, the ``paradox'' arises because logicians have always tried to formalize self-reference within structural logics, which are not adequate to describe such a strong logical correlation, which resembles quantum entanglement. However, the appropriate connective was missing,
until we found it in a recent paper \cite{Zi06} and called it $@
=$ ``entanglement''.
The introduction of the connective $@$ is possible only in a substructural logic obtained by reflection of a metalanguage that is the mathematical formalism of quantum computing. In \cite{Zi06} we showed that such a logic is Basic logic \cite{SBF00}.

Basic logic can accommodate the connective for quantum
superposition ($\&$ =``and'') and the new connective for quantum
entanglement ($@ =$ ``entanglement'').

The connective $\&$ in Basic logic can describe in fact quantum
superposition because Basic logic is a paraconsistent logic
\cite{Pri02}.

The definition of $@$ is possible because Basic logic is a
substructural logic (it has neither the contraction rule, nor the
weakening rule), and is non-distributive.

As quantum superposition together with quantum entanglement lead
to massive quantum parallelism that is the source of quantum
computational speed-up \cite{JoLi??}, we argued \cite{Zi06} that Basic logic
should be the most adequate logic for quantum computing. Also, we
noticed that the absences of the contraction rule and of the
weakening rule correspond, in quantum computing, to the no-cloning
\cite{WoZu82} and no-erase \cite{PaBra00} theorems respectively.

Among the properties of the connective $@$, we found that $@$ is
non- idempotent. The non-idempotence of $@$ is strictly related to
the physical fact that self-entanglement is meaningless, unless
one could clone the original qubit, which is impossible because of
the no-cloning theorem. And as we already mentioned, the quantum
no-cloning theorem reflects itself into logic as the absence of
the contraction rule (data cannot be copied).

Although at a first sight the non-idempotence of $@$ might seem a
quite harmless property, in fact it is a metatheorem with
intriguing consequences.

In this paper we prove the theorem, and explore its consequences.
The theorem states that a self-entangled sentence cannot simply
exist, as if it did, it would lose its identity. This becomes
apparent when the metalanguage is reflected into a substructural
logic like Basic logic, once the latter is equipped with the
connective $@$.

On the other hand, it is a fact that self-referential sentences and
paradoxes do
appear in natural language, but
this happens when the latter is the only metalanguage (with no hierarchy) and is reflected into a structural logic,
or more generally, into a logic which cannot include the
connective $@$.

Section \ref{Sec2} is a short review of some results of
Ref.~\cite{Zi06}, namely, the definitions of quantum superposition
and quantum entanglement in logical terms, and, in particular, the
definition of the connective $@$.

Section \ref{Sec3} consists of the statement and interpretation
of the no-self-reference metatheorem of Basic logic, which is
based on the property of non-idempotence of the connective
$@$. Then, we relate our results to the ``Liar paradox''.

In Section \ref{Sec4}, we introduce a new, generalized definition
of self-reference, which can deal with substructural logics, and
which reduces to the standard one in the case of structural logics.

In the Appendix A, we give the formal proof of the metatheorem discussed
in Section \ref{Sec3}.

\section{The Logical Connective for Quantum Entanglement}
\label{Sec2}

The qubit is the unit of quantum information \cite{NiChu00}:

\begin{equation}
\label{eq1} |Q\rangle=a|0\rangle+b|1\rangle
\end{equation}
where $\{|0\rangle,|1\rangle\}$ is called the computational basis
and $a$, $b$ are complex numbers called probability amplitudes
such that $|a|^2+|b|^2=1$.
Two qubits are said entangled when the bipartite state
$|Q\rangle_{AB}$ is not-separable, i.e.
$$
|Q\rangle_{AB}\neq |Q\rangle_A\otimes |Q\rangle_B\enspace,
$$

where $\otimes$ is the tensor product in Hilbert spaces. When the
composite system of two qubits is in a non-separable state, it is
impossible to attribute to each qubit a pure state, as their
states are superposed with one another.

In particular, a bipartite state of two qubits is maximally
entangled when it is one of the four Bell's states \cite{Be87}:

\begin{equation}
\begin{split}
\label{eq2} |\Phi_\pm\rangle_{AB}& =
\frac{1}{\sqrt{2}}\left(|0\rangle_A \otimes |0\rangle_B \pm
|1\rangle_A\otimes|1\rangle_B\right)\enspace,\\
|\Psi_\pm\rangle_{AB} & =\frac{1}{\sqrt{2}}
\left(|0\rangle_A\otimes|1\rangle_B\pm|1\rangle_A\otimes|0\rangle_B\right)\enspace.
\end{split}
\end{equation}

For simplicity, in this paper we will consider only Bell states.

As we showed in \cite{Zi06}, expressing the qubit $|Q\rangle_A$ in
\eqref{eq1} in logical terms leads to the compound proposition:

\begin{equation}
\label{eq3} Q_A\doteq A\& A^\perp
\end{equation}
where the atomic proposition $A$ is associated with bit
$|1\rangle$, its primitive negation $A^\perp$ is associated with
bit $|0\rangle$, and the (right) connective $\&$ =``and'' is the
additive conjunction \cite{SBF00}.

In the same way, the second qubit $|Q_B\rangle$  is expressed, in
logical terms, by a second compound proposition $Q_B\doteq B\&
B^\perp$.

Bell's states will be expressed, in logical terms, by the
expression

\begin{equation}
\label{eq4} Q_{Bell}=Q_A@Q_B
\end{equation}
where $@$ is the new logical connective called ``entanglement'' \cite{Zi06}.

The logical structure for, say, the
Bell's states $|\Phi_\pm\rangle_{AB}$  (which was given in \cite{Zi06} as
the definition of $@$) is:

\begin{equation}
\label{eq5} Q_A@Q_B \doteq (A\wp B)\& (A^\perp\wp B^\perp)
\end{equation}
where the (right) connective $\wp$ =``par'', introduced by Girard
in Linear logic \cite{Gi87} is the multiplicative disjunction, the
dual of the (left) connective $\otimes$ = ``times'', which is the
multiplicative conjunction.
The logical rules for $@$, as well as its properties, can be found in \cite{Zi06}.
In what follows, we will discuss the non-idempotence of @, also in relation with self-reference, and with the Liar paradox.

\section{Non-idempotence of $@$ and the Liar Paradox}
\label{Sec3}

Here we give an informal proof of the non-idempotence of $@$ (a
more formal proof will be given in the Appendix A).
We want to prove:

\begin{equation}
\label{eq6}
Q_A@Q_A\neq Q_A\enspace.
\end{equation}
From the definition of $@$ in \eqref{eq5}, by replacing $B$ with
$A$ we get:

\begin{equation}
\label{eq7} Q_A@ Q_A \doteq (A\wp A)\& (A^\perp\wp A^\perp)\neq
Q_A
\end{equation}
as $\wp$  is non-idempotent. In fact, to prove the idempotence of
$\wp$ would require the validity of both the contraction and
weakening rules.

If one makes the formal proof one has to go both ways:  to show
that $A\vdash A\wp A$ does not hold because of the absence of the
weakening rule, and that $A\wp A\vdash A$  does not hold because
of the absence of the contraction rule.

If instead weakening and contraction did hold, then
$\wp\equiv\vee$ ($\otimes\equiv\&$), and from the definition of
$@$  we would get:

\begin{equation}
\label{eq8}
Q_A@Q_A\doteq(A\vee A)\&(A^\perp\vee A^\perp)=Q_A
\end{equation}
because of the idempotence of $\vee$.
In that case, from the definition of the dual $\S$ of $@$ given in \cite{Zi06}, namely:

\begin{equation}
\label{eq9}
Q_A \S Q_B \doteq (A\otimes B) \vee (A^\perp \otimes B^\perp)
\end{equation}
we will
also get:

\begin{equation}
\label{eq10}
Q_A\S Q_A\doteq(A\& A)\vee(A^\perp\& A^\perp)=Q_A
\end{equation}
because of the idempotence of $\&$.

Notice, in particular, that the formal proof that the dual of
$\wp$, namely $\otimes$ = ``times'' is non-idempotent, would
exchange the roles of the contraction and the weakening rules used
in the proof done for $\wp$.
The fact that $\otimes$ is non-idempotent, leads to the result
that the dual of $@$, namely $\S$, is non-idempotent either. Then,
$@$ ($\S$) is non-idempotent because $\wp$($\otimes$) is
non-idempotent.

This illustrates an obvious physical fact: self-entanglement
(entanglement of a qubit with itself) is impossible as it would
require a quantum clone, which is forbidden by the no-cloning
theorem. In a sense, one can say that the two main no-go theorems
of quantum computing, namely the no-cloning and no-erase theorems
are (logically) dual to each other. And the no-self-entanglement
``corollary'' is a consequence of the first one, when entanglement
is expressed in terms of $\S$, and a consequence of the second
one, when entanglement is expressed by $@$.

On the other hand, it turns out that the meaning of ``no-self-entanglement'' is much more profound in logic. In fact,
affirming that in a certain formal language it is impossible to
get a (compound) proposition (maximally) entangled with itself
means that the language under study does not lead to
self-referential sentences in the corresponding metalanguage.
Schematically:

\begin{tabular}{ccccc}
&&&&\\
\multicolumn{5}{c}{\textbf{BASIC LOGIC}}\\
\footnotesize\textbf{No contraction} & \footnotesize\textbf{No
weakening} & $\longleftrightarrow$ & \footnotesize\textbf{No
contraction} & \footnotesize\textbf{No
weakening}\\
$\downarrow$ & $\downarrow$ & \footnotesize\textbf{Symmetry} &
$\downarrow$ & $\downarrow$\\
\footnotesize $A\vdash A\otimes A$&
 \footnotesize  $ A\otimes A\vdash A$ &
$\footnotesize\otimes\longleftrightarrow\wp$ & \footnotesize $A\wp
A\vdash A$ & \footnotesize $A\vdash A\wp A$\\
\multicolumn{2}{c}{\footnotesize Cannot be proved} & &
\multicolumn{2}{c}{\footnotesize Cannot be proved}\\
\multicolumn{2}{c}{$\downarrow$} &
\scriptsize\textbf{No-idempotence} &
\multicolumn{2}{c}{$\downarrow$}\\
\multicolumn{2}{c}{\footnotesize$Q_A\S Q_A\neq Q_A$} &
\textbf{\footnotesize of $\S$ and @} &\multicolumn{2}{c}{\footnotesize$Q_A@Q_A\neq Q_A$}\\
&&&&\\
\multicolumn{5}{c}{\textbf{QUANTUM COMPUTING}}\\
\multicolumn{2}{c}{\footnotesize\textbf{No-cloning}} &
$\longleftrightarrow$ &
\multicolumn{2}{c}{\footnotesize\textbf{No-erase}}\\
\multicolumn{2}{c}{\footnotesize$|\Psi\rangle\nrightarrow|\Psi\rangle
\otimes |\Psi\rangle$} & $\downarrow$ &
\multicolumn{2}{c}{\footnotesize$|\Psi\rangle\otimes|\Psi\rangle
\nrightarrow
|\Psi\rangle$}\\
& \multicolumn{3}{c}{\footnotesize\textbf{No self-entanglement}} &\\
&&&&\\
\end{tabular}

Notice that the property of the non-idempotence of $@$ (and of its
dual $\S$) deals with the object language. However, when this
property is translated into a metalanguage like natural language,
we get:

$$
\underbrace{Q_A}_{\text{\emph{a sentence}}}
\underbrace{@}_{\text{\emph{entangled with}}}
\underbrace{Q_A}_{\text{\emph{itself}}}
\underbrace{\neq}_{\text{\emph{is not}}}
\underbrace{Q_A}_{\text{\emph{itself}}}
$$

Then we state the following.

\begin{mththm}
\label{meta1}
A sentence $Q_A$ logically entangled with itself
is not itself.
\end{mththm}

\begin{figure}
\centering \psfrag{Object languages}{\textbf{Object languages}}
\psfrag{Metalanguages}{\textbf{Metalanguages}}
\psfrag{Q@Q}{$Q@Q\neq Q$}
\includegraphics[height=7cm]{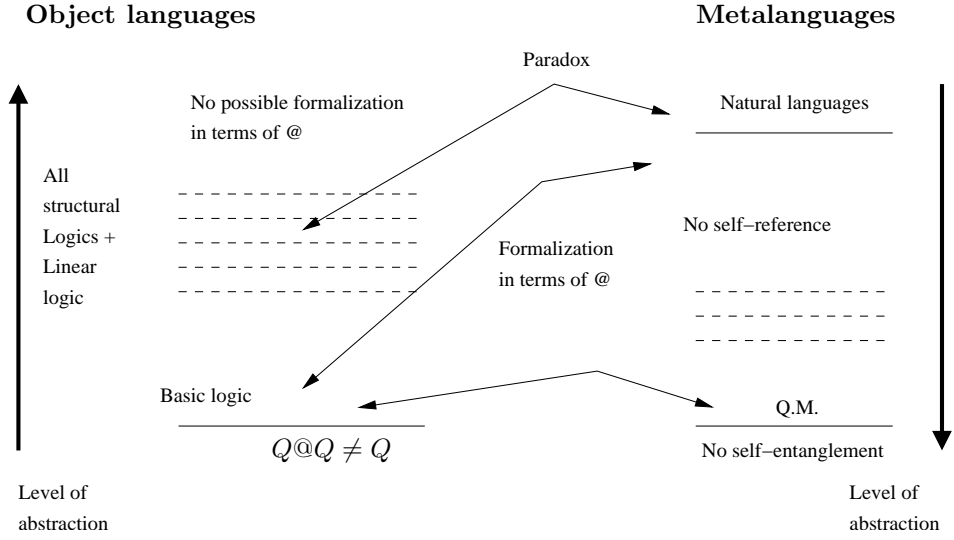}
\caption{The Liar ``Paradox'' Revisited} \label{Fig1}
\end{figure}

This just expresses the impossibility of having self-entangled
sentences in the metalanguage. It is not a paradox. However, many
classical ``paradoxes'' like the Liar paradox: ``This sentence is
false'' look very much like the property of the non-idempotence of
$@$, which instead is a metatheorem. The reason is that in our
classical reasoning, the concept of entanglement is missing. Moreover, when we try to formalize the ``paradox'' in any other
logic (but Basic logic) which is lacking in the connective $@$, we fail.
See Fig.\ref{Fig1}.

\section{A generalized definition of self-reference }
\label{Sec4}

Self-reference, as usually understood, can be interpreted as a function $F$ from the object-language $L_0$ to the metalanguage $L_m$, that is:

\begin{equation}
\label{eq11}
F\colon L_0 \to L_m
\end{equation}
such that

\begin{equation}
\label{eq12}
S:\equiv F(\text{``}S\text{''})
\end{equation}
where $S\in L_m$ is a sentence in the metalanguage, and ``$S$''$\in L_0$ , is the name of  $S$ in the object-language.
A self-referential sentence of kind \eqref{eq12} is, for example, the Liar sentence $L$: ``This sentence is false'', i.e.

\begin{equation}
\label{eq13}
L:\equiv \neg True (\text{``}L\text{''}) \qquad (\text{with: } F\equiv \neg True)
\end{equation}
By the use of Tarski's definition of truth \cite{Ta82}:

\begin{equation}
\label{eq14}
True (\text{``}A\text{''})\equiv A
\end{equation}
Which is an axiom schema for an arbitrary formula $A$, the Liar sentence \eqref{eq13} becomes:

$$
L:\equiv \neg True (\text{``}L\text{''})\equiv \neg L
$$
leading to the self-referential paradox:

\begin{equation}
\label{eq15}
L\equiv\neg L
\end{equation}
Here, we introduce a more general definition of self-reference:

\begin{equation}
\label{eq16}
S:\equiv F(f(\text{``}S\text{''}))
\end{equation}
Where $f$ is a function from the object-language into itself:

$$
f\colon L_0\to L_0
$$
The function $f$ is built as follows. Let us consider the name of $S$, i.e., ``$S$'' in the object-language (``$S$''$\in L_0$), let us duplicate it (``$S$'', ``$S$'') by the ``diagonal'' function $\Delta$:

$$
\Delta\colon L_0 \to L_0\times L_0
$$
If the name of $S$ is a formula of $L_0$, (let us call $S_0$) we can send the pair $(S_0,S_0)$  into a new formula $S_0\bullet S_0$ by the binary connective $\bullet$

$$
(S_0,S_0) \stackrel{\bullet}{\rightarrow} S_0\bullet S_0
$$
Then, $f$ is the composite function $f=\bullet\circ\Delta$

$$
f\colon L_0 \stackrel{\Delta}{\rightarrow}L_0\times L_0 \stackrel{\bullet}{\rightarrow}
L_0
$$

There are two cases:

\begin{enumerate}

\item
The connective $\bullet$ is idempotent (for example: $\bullet\equiv\&=\text{``$and$''}$, $\bullet \equiv\vee= \text{``$or$''}$):

$$
S_0\bullet S_0=S_0
$$
That is, the function $f$ has a fixed point:

\begin{equation}
\label{eq17}
f(\text{``}S\text{''})=\text{``}S\text{''}
\end{equation}
In this case, \eqref{eq16}, by the use of \eqref{eq17}, reduces to
\eqref{eq12}, namely, to the standard definition of
self-reference.

\item
The connective $\bullet$ is non-idempotent (for example: $\bullet\equiv\otimes=\text{``$times$''}$, $\bullet\equiv\wp=\text{``$par$''}$, $\bullet\equiv @=\text{``$entanglement$''}$):

$$
S_0\bullet S_0 \neq S_0
$$
That is, the function $f$ has no fixed point:

\begin{equation}
\label{eq18}
f(\text{``}S\text{''}) = \text{``}\widehat{S}\text{''}\neq \text{``}S\text{''}
\end{equation}
In this case, \eqref{eq16} cannot be rewritten as \eqref{eq12}.
Notice that this fact is peculiar of sub-structural logics,
like Basic logic and linear logic, which can accommodate
multiplicative connectives like $\otimes$, $\wp$ and $@$ (more
precisely, Basic logic can accommodate all the three of them, and
linear logic only two of them, namely $\otimes$ and $\wp$). In
structural logics like quantum logic, intuitionistic logic, and
classical logic, $\otimes$ ``collapses'' to $\&$, which is
idempotent, $\wp$  ``collapses'' to $\vee$, which is also
idempotent, and $@$ ``collapses'' to a function of $(\vee,\&)$,
which is idempotent as well.
\end{enumerate}

In summary, in sub-structural logics, one is allowed to adopt the
generalized version \eqref{eq16} of self- reference, while in
structural logics, \eqref{eq16} reduces to 
\eqref{eq12}, and one has to stick with the standard definition of
self-reference.

In case 2.,  \eqref{eq16}, by the use of \eqref{eq18}, becomes:

\begin{equation}
\label{eq19}
S:\equiv F(\text{``}\widehat{S}\text{''})
\end{equation}
In particular, if $S$ is the Liar sentence $L$, and $F\equiv \neg
True$, we get, from  \eqref{eq19}:

\begin{equation}
\label{eq20} L:\equiv\neg True (\text{``}\widehat{L}\text{''})
\qquad (\text{with: } \text{``}\widehat{L}\text{''} \neq
\text{``}\widehat{L}\text{''})
\end{equation}
Equation \eqref{eq20}, by the use of  Tarski's definition of truth
\eqref{eq14} becomes:

\begin{equation}
\label{eq21} L:\equiv \neg True (\text{``}\widehat{L}\text{''})
\equiv \neg \widehat{L}
\end{equation}                                                                                  
hence:

\begin{equation}
\label{eq22} L= \neg \widehat{L}
\end{equation}
which is not a paradox.

Let us consider now the composite function $F(f(''S''))$ which
appears in the generalized definition of self-reference
\eqref{eq16}, and let us call it $\sigma$.

\begin{equation}
\label{eq23} \sigma\colon L_0\to L_m
\end{equation}
See the following diagram:

$$
\xymatrix{
L_0 \ar[d]_{F}&\\
L_m & L_0 \ar[l]^{\sigma} \ar[ul]_{f}
}
$$


\begin{equation}
\label{eq24} f\circ F=\sigma
\end{equation}

If the connective $\bullet$ is idempotent, the interpretation of
$(f\circ F)(''S'')$ is $S$ in the metalanguage, and we get:

\begin{equation}
\label{eq25} F\equiv\sigma
\end{equation}

There are only two possible cases of study: linear logic with $\otimes$
(it is sufficient to consider only $\otimes$, as $\wp$  is the dual of
$\otimes$), and Basic logic, with both $\otimes$ and $@$.

The connective $\otimes$ in the object-language is the ``reflection'' 
of the (physical) tensor product $\otimes$  (where the same symbol is used) 
in the quantum mechanical metalanguage. 

The fact that a multiplicative connective  $\bullet$ in the object-language 
is non-idempotent, is a necessary but not sufficient condition for having a 
generalized version of self-reference.
The sufficient condition requires that the corresponding (physical) link in 
the metalanguage does never allow cloning.
This is not the case of the tensor product  $\otimes$, because one can clone a 
basis state:

\begin{equation}
\label{eq26}
|0\rangle \to |0\rangle \otimes |0\rangle\enspace, \qquad 
|1\rangle \to |1\rangle \otimes |1\rangle
\end{equation}

by means of the reversible XOR, or controlled NOT (CNOT) 
quantum logic gate. The CNOT gate operates on two qubits $a$ and $b$.
See the following diagram:

\begin{equation}
\label{Fig3}
\xymatrix{
a& \bullet \ar@{-}[l]_{\text{``control''}} \ar@{-}[r] \ar@{-}[d] & a\phantom{\otimes b}\\
b & \bigoplus \ar@{-}[l]_{\text{``target''}} \ar@{-}[r] & a\oplus b
}
\end{equation}

The $\oplus$ symbol in the diagram \eqref{Fig3} represents modulo $2$ addition.
The CNOT flips the ``target'' input if its ``control'' input is $|1\rangle$, 
and does nothing if it is $|0\rangle$. 
So, one can clone a basis state, in the two cases:

\begin{equation*}
\begin{cases}
control \, |0\rangle\\
target \, |0\rangle
\end{cases}
\xrightarrow{XOR} |00\rangle
\qquad
\begin{cases}
control \, |1\rangle\\
target \, |0\rangle
\end{cases}
\xrightarrow{XOR} |11\rangle
\end{equation*}
(where $|00\rangle$ and $|11\rangle$ stand for the tensor products 
$|0\rangle\otimes|0\rangle$  and $|1\rangle\otimes|1\rangle$  respectively).

We recall that the basis states are, in the corresponding object-language, 
the atomic formulas $A=|1\rangle$, $A^\perp =|0\rangle$. Then, \eqref{eq26} does not hold in the object-language, as the connective $\otimes$=``$times$''  is non- idempotent:

$$
A\otimes A\neq A \enspace, \qquad A^\perp\otimes A^\perp\neq A^\perp \enspace.
$$

But when one tries to clone a superposed state, for example the ``cat'' state:

$$
|Q\rangle_{Cat} = \frac{1}{\sqrt{2}} (|0\rangle+|1\rangle)
$$
by means of the CNOT, one gets a Bell state, which is maximally entangled:

\begin{equation*}
\begin{cases}
control\, \frac{1}{\sqrt{2}} (|0\rangle+|1\rangle)\\
target |0\rangle
\end{cases}
\xrightarrow{XOR}
\frac{1}{\sqrt{2}} (|00\rangle+|11\rangle)
\end{equation*}
In fact, by the no-cloning theorem, it is forbidden to copy an unknown quantum state:

\begin{equation}
\label{eq27}
|Q\rangle \nrightarrow |Q\rangle\otimes|Q\rangle
\end{equation}
Equation \eqref{eq27} holds in the object language as well:

$$
Q\otimes Q\neq Q
$$

Nevertheless, the problem with the connective ``times'' does appear in the case of basis states, as the corresponding physical link, the tensor product, can form a clone in that case, and this is enough to invalidate the sufficient condition.
 
Let us consider now the sufficient condition in more detail.
In the schema of the CNOT, in Diagram \eqref{Fig3}, we see that we have cloning for $b = 0$. Let us call cloning the function:

$$
Clon\colon (a,0)\to (a,a)
$$
In the case with $a = 1$:
$$
Clon\equiv (id, NOT)\colon(1,0)\to(1,1)
$$
In the case with $a = 0$:
$$
Clon\equiv(id,id)\colon (0,0)\to (0,0)
$$
Let us consider the case with $a = 1$. See the following diagram

\begin{equation}
\label{Fig4}
\xymatrix{
L_{m\;\;(1,0)} \ar[dr]^{\;\;\mathrm{Clon}\equiv(\mathrm{id},\mathrm{NOT})}&\\
L_{0\;\;(A,A^\perp)} \ar[u]_{\sigma} \ar[r]^{(F,F')} & L_m\times L_{m\;\;(1,1)}
}
\end{equation}

With:

\begin{equation*}
Clon\colon L_m\to L_m\times L_m\enspace, \quad
Clon\equiv(id,NOT)\equiv
\begin{cases}
id\colon 1\to 1\\
NOT\colon 0\to1
\end{cases}
\end{equation*}

\begin{equation*}
\sigma\colon L_0\to L_m\enspace, \qquad \sigma\colon 
\begin{cases}
A\to 1\\
A^\perp\to 0
\end{cases}
\end{equation*}

$$
(F,F')\colon L_0\to L_m\times L_m 
$$

\begin{equation*}
\begin{split}
F\colon & A\to 1\\
F'\colon & Aì\perp \stackrel{\perp}{\rightarrow} A \stackrel{F}{\rightarrow}1
\end{split}
\quad F'=F\circ \perp
\end{equation*}

\begin{equation}
\label{eq28}
\begin{cases}
\sigma\circ id =F\\
\sigma\circ NOT = F\circ \perp
\end{cases}
\Rightarrow F=\sigma
\end{equation}

The case with $a = 0$ is very similar, and is left as an exercise for the willing reader.
Equation \eqref{eq28} says that if it exists a function $Clon$ in the metalanguage, such that the diagram \eqref{Fig4} commutes, we recover the usual definition of self-reference, even if the corresponding logical connective in the object-language (in this case ``times'') is non-idempotent, and we cannot skip self-referential paradoxes.
Then, to be sure that a generalized version of self-reference is present in a given logic, one must check not only that the connective  in the object-language is non-idempotent, but also that the corresponding physical link in the metalanguage (in this case the tensor product) does not lead to cloning in some particular case. In a sense, the ``reflection'' of the metalanguage into the object-language is in general incomplete, as far as self-reference is concerned.
This is not the case, however, for the physical link of entanglement, because entangling a basis state with itself is physically meaningless (entanglement is a particular kind of superposition).
On the other hand, as we have already seen, entangling an (unknown) quantum state with itself  is forbidden by the no-cloning theorem.
Then, in the case of entanglement, the reflection between the metalanguage and the object language is complete, and one can adopt the generalized version of self-reference. 

Our conclusion is that Basic logic is the unique formal language that can reflect a metalanguage, like that of a quantum computer in a entangled state, which is never self-referential (and hence is completely paradox-free). Differently stated, the halting problem in quantum computing appears meaningless once the adequate logical language (Basic logic) is adopted.

\section*{Appendix A}

In this appendix we give a formal proof of Metatheorem \ref{meta1}
discussed in Section \ref{Sec3}. We try proving $Q_A@Q_A\doteq
Q_A$.

Let us try first $Q_A@Q_A\vdash Q_A$.

There are no rules of Basic logic that we can use in the
derivation, which can lead to a proof:

$$
\frac{}{Q_A@Q_A\vdash Q_A}
$$

And, as the cut-elimination theorem holds in Basic logic
\cite{FaSa98}, we are sure that there are no other rules leading
to a proof.

Let us try now the other way around: $Q_A\vdash Q_A@Q_A$.

The only rule we can use in the derivation is the $@$-formation
rule, and there are no further rules in Basic logic, which would
lead to a proof:

$$
\frac{}{\displaystyle\frac{Q_A\vdash Q_A,Q_A}{Q_A\vdash
Q_A@Q_A}\,@-form.}
$$

And, again, because of cut-elimination, we are sure that there are
no other rules leading to a proof. For the sake of the physical
interpretation, we show now that in the case the contraction and
weakening rules did hold, the proof would be possible.

Let us prove first $Q_A@Q_A\vdash Q_A$.

$$
\frac{Q_A\vdash Q_A \qquad Q_A\vdash Q_A}{\displaystyle
\frac{Q_A@Q_A\vdash Q_A,Q_A}{Q_A@Q_A\vdash
Q_A}\,contr.}\,@-\mathrm{exp}l.refl.
$$

Let us prove now the other way around $Q_A\vdash Q_A@Q_A$.

$$
\frac{Q_A\vdash Q_A}{\displaystyle \frac{Q_A\vdash
Q_A,Q_A}{Q_A\vdash Q_A@Q_A}\,@-form.}\,weak.
$$

It is impossible to prove $Q_A\vdash Q_A @ Q_A$ in Basic logic,
because the weakening rule (in the step \emph{weak.}) does not
hold. In conclusion, it is impossible to prove idempotence of $@$
in Basic logic, because of the absence of the two structural rules
of weakening and contraction.

\bibliographystyle{unsrt}
\bibliography{liar_paradox}

\begin{thebibliography}{10}

\bibitem{Zi06}
P.~A. Zizzi.
\newblock Basic logic and quantum entanglement.
\newblock arXiv: quant-ph/0611119, submitted for publication in the Proceedings
  of DICE 2006 (Journal of Physics: Conference Series, Publ. Institute of
  Physics, London).

\bibitem{SBF00}
G.~Sambin, G.~Battilotti, and C.~Faggian.
\newblock Basic logic: reflection, symmetry, visibility.
\newblock {\em J. Symbolic Logic}, 65(3):979--1013, 2000.

\bibitem{Pri02}
G.~Priest.
\newblock Paraconsistent logic.
\newblock In {\em Handbook of philosophical logic (Second Edition), Vol.\ 6},
  pages 287--393. Kluwer Acad. Publ., Dordrecht, 2002.

\bibitem{JoLi??}
R.~Josza and N.~Linden.
\newblock On the role of entanglement in quantum computational speed-up.
\newblock arXiv: quant-ph/0201143.

\bibitem{WoZu82}
W.~K. Wootters and W.~H. Zureck.
\newblock A single quantum cannot be cloned.
\newblock {\em Nature}, 299:802, 1982.

\bibitem{PaBra00}
A.~K. Pati and Braunstein~S. L.
\newblock Impossibility of deleting an unknown quantum state.
\newblock {\em Nature}, 404:164, 2000.
\newblock arXiv: quant-ph/9911090.

\bibitem{NiChu00}
M.~A. Nielsen and I.~L. Chuang.
\newblock {\em Quantum computation and quantum information}.
\newblock Cambridge University Press, Cambridge, 2000.

\bibitem{Be87}
J.~S. Bell.
\newblock {\em Speakable and unspeakable in quantum mechanics}.
\newblock Cambridge University Press, Cambridge, 1987.
\newblock Collected papers on quantum philosophy.

\bibitem{Gi87}
J.~Y. Girard.
\newblock Linear logic.
\newblock {\em Theoret. Comput. Sci.}, 50(1):1--102, 1987.

\bibitem{Ta82}
A.~Tarski.
\newblock {\em Logic, semantics, metamathematics}.
\newblock Hackett Publishing Co., Indianapolis, second edition, 1983.
\newblock Papers from 1923 to 1938, Translated by J. H. Woodger, Edited and
  with an introduction by John Corcoran.

\bibitem{FaSa98}
C.~Faggian and G.~Sambin.
\newblock From basic logic to quantum logics with cut-elimination.
\newblock In {\em Proceedings of the International Quantum Structures
  Association 1996 (Berlin)}, volume~37, pages 31--37, 1998.

\end{thebibliography}

\end{document}